\documentclass{svjour2}

\smartqed

\usepackage{graphicx}

\journalname{Int J Theor Phys}

\begin{document}

\title{Phenomenology of $\Lambda$-CDM model: a
possibility of accelerating Universe with positive pressure}

\titlerunning{Phenomenology of $\Lambda$-CDM model ...}

\author{Saibal Ray \and Maxim Khlopov \and Utpal Mukhopadhyay \and Partha Pratim
Ghosh}

\authorrunning{Mukhopadhyay \and Ghosh \and Khlopov \and Ray}

\institute{Saibal Ray \at Department of Physics, Government
College of Engineering and Ceramic Technology, Kolkata 700 010,
West Bengal, India \email{saibal@iucaa.ernet.in} \and Maxim
Khlopov \at Center for Cosmoparticle physics ``Cosmion", 125047,
Moscow, Russia\\ National Research Nuclear University ``Moscow
Engineering Physics Institute", 115409 Moscow, Russia \&
\\APC laboratory 10, rue Alice Domon et L{\'e}onie Duquet 75205
Paris Cedex 13, France \email{khlopov@apc.univ-paris7.fr} \and
Utpal Mukhopadhyay \at Satyabharati Vidyapith, Nabapalli, North 24
Parganas, Kolkata 700 126, West Bengal, India
\email{utpal1739@gmail.com} \and Partha Pratim Ghosh \at
Department of Physics, A. J. C. Bose Polytechnic, Berachampa,
North 24 Parganas, West Bengal, India
\email{parthapapai@gmail.com}}

\date{Received: date / Accepted: date}

\maketitle

\begin{abstract}
Among various phenomenological $\Lambda$ models, a time-dependent
model $\dot \Lambda\sim H^3$ is selected here to investigate the
$\Lambda$-CDM cosmology. The model can follow from dynamics,
underlying the origin of $\Lambda$. Using this model the
expressions for the time-dependent equation of state parameter
$\omega$ and other physical parameters are derived. It is shown
that in $H^3$ model accelerated expansion of the Universe takes
place at negative energy density, but with a positive pressure. It
has also been possible to obtain the change of sign of the
deceleration parameter $q$ during cosmic evolution.


\end{abstract}

\section{Introduction}
Any theoretical model related to cosmology should be supported by
observational data. Present cosmological models suggest that the
Universe is primarily made of dark matter and dark energy. Various
observational evidences, including SN
Ia~\cite{Riess1998,Perlmutter1999,Knop2003,Riess2004,Tegmark2004a,Astier2005,Spergel2007}
data, support the idea of accelerating Universe and it is supposed
that dark energy is responsible for this effect of speeding up.
Now-a-days it is accepted that about two third of the total energy
density of the Universe is dark energy and the remaining one third
consists of visible matter and dark matter \cite{Sahni2004}.

Though dark matter had a significant role during structure
formation in the early Universe, its composition is still unknown.
It is predicted that the dark matter should be non-baryonic and
various particle physics candidates and their mixtures are
discussed (see e.g. \cite{khlopov95,khlopovbook1} for review). In
particular, time-varying forms of dark matter
\cite{dorr1,Turner:1984nf,Gelmini:1984pe} in Unstable Dark Matter
(UDM) scenarios \cite{dorr1,dorr2,dorr3} are still not fully
explored and deserve interest, giving simultaneously clustered and
unclustered dark matter components.

The standard cold dark matter (SCDM) model introduced in 1980's
which assume $\Omega_{CDM}=1$ is out of favor today
\cite{Efstathiou1990,Pope2004}. After the emergence of the idea of
accelerating Universe, the SCDM model is replaced by $\Lambda$-CDM
(or LCDM) model. This model includes dark energy as a part of the
total energy density of the Universe and is in nice agreement with
various sets of observations \cite{Tegmark2004b}. In this
connection it is to be noted here that according to $\Lambda$-CDM
model, acceleration of the Universe should be a recent phenomenon.
Some recent works \cite{Padmanabhan2003,Amendola2003} favor the
idea that the present accelerating Universe was preceded by a
decelerating one and observational evidence \cite{Riess2001} also
support this.

 Now, in most of the recent cosmological research, the equation of state parameter
$\omega$ has been taken as a constant. However, its seems that for
better result $\omega$ should be taken as time-dependent
 \cite{Chervon2000,Zhuravlev2001,Peebles2003}. Therefore, in
 the present work $\Lambda$-CDM Universe has been investigated by
 selecting a specific time-dependent form of $\Lambda$, viz., $\dot \Lambda\sim
 H^3$ along with $\omega(t)$. A kind of this $\Lambda$-model was previously studied by
Reuter and Wetterich \cite{Reuter1987} for finding out an
explanation of the presently observed small value of $\Lambda$.
This choice found realization in the approach treating
time-variation of $\Lambda$ as Bose condensate
evaporation~\cite{Dymnikova:2001ga} in the framework of model of
self consistent
inflation~\cite{Dymnikova:2001ga,Dymnikova:2001jy,Dymnikova:1998ps}.
Very recently it is also used by Mukhopadhyay et al.
\cite{Mukhopadhyay2007} for investigating various inherent
features of the $\Lambda$-CDM Universe.

In this context it is important to note that effects of real and
virtual particles can play nontrivial role in the origin of
$\Lambda$ and its time-variation. Indeed, $R^2$ term was induced
by vacuum polarisation and gave rise to inflation in one of the
first inflationary models~\cite{Starobinsky:1980te}. The back
reaction of particle products of Bose-Einstein condensate
evaporation slowed down evaporation and damped coherent scalar
field oscillations in the approach of self consistent
inflation~\cite{Dymnikova:2001ga,Dymnikova:2001jy,Dymnikova:1998ps}.
It resulted in specific time dependence of $\Lambda$, being
determined by the cross section of interaction of evaporated
particle. The maximal estimation of this cross section gave the
time variation of
$\Lambda$~\cite{Dymnikova:2001ga,Dymnikova:2001jy}, leading to
possible realization of our model in the form $\dot \Lambda\sim
H^3$.

 Therefore, motivated by the
time variation of $\Lambda$ and using the phenomenological model
of the kind $\dot \Lambda\sim H^3$ the expression for the
time-dependent equation of state parameter $\omega$ and various
physical parameters are derived in the present investigation. The
change in sign of the deceleration parameter q has also been shown
in this case which refers to the evolution of the Universe via a
phase transition from deceleration to the present acceleration.

Interestingly the expression for energy density obtained in the
present model is negative. However, this type of negative density
is not at all unavailable in the literature. It can be mentioned
here that the negative energy density was first obtained by
Casimir \cite{Casimir1948}. Hawking found the existence of
negative energy density at the horizon of a black hole
\cite{Hawking1975}. Davies and Fulling also studied about the
negative energy fluxes in the radiation from moving mirrors
\cite{Davies1976,Fulling1977}.

The scheme of the present investigation is as follows: in Section
2 the field equations are provided whereas their solutions as well
as the physical features have been sought for in the Sections 3
and 4. In the Section 5 some concluding remarks have been made.

\section{Field Equations}
The Einstein field equations are given by
\begin{eqnarray}
R^{ij}-\frac{1}{2}Rg^{ij}= -8\pi G\left[T^{ij}-\frac{\Lambda}{8\pi
G}g^{ij}\right]
\end{eqnarray}
where the cosmological term $\Lambda$ is time-dependent, i.e.
$\Lambda = \Lambda(t)$ and $c$, the velocity of light in vacuum is
assumed to be unity.

Let us consider the Robertson-Walker metric
\begin{eqnarray}
ds^2=-dt^2+a(t)^2\left[\frac{dr^2}{1-kr^2}+r^2(d\theta^2+sin^2\theta
d\phi^2)\right]
\end{eqnarray}
where $k$, the curvature constant, assumes the values $-1$, $0$
and $+1$ for open, flat and closed models of the Universe
respectively and $a=a(t)$ is the scale factor. For the spherically
symmetric metric (2), the field equations (1) yield respectively
the Friedmann and Raychaudhuri equations which can be given by
\begin{eqnarray}
3H^2+\frac{3k}{a^2}= 8\pi G\rho+\Lambda,
\end{eqnarray}
\begin{eqnarray}
3H^2+3\dot H= -4\pi G(\rho+3p)+\Lambda
\end{eqnarray}
where $G$, $\rho$ and $p$ are the gravitational constant,
matter-energy density and fluid pressure respectively. Here the
Hubble parameter $H$ is related to the scale factor $a$ by the
relation $H=\dot a/a$. In the present work, $G$ is assumed to be
constant. The generalized energy conservation law for variable $G$
and $\Lambda$ is derived by Shapiro et al. \cite{Shapiro2005}
using Renormalization Group Theory and also by Vereschagin et al.
\cite{Vereschagin2006} using a formula of Gurzadyan and Xue
\cite{Gurzadyan2003}.

The conservation equation for variable $\Lambda$ and constant $G$
is a special case of the above mentioned generalized conservation
law and is given by
\begin{eqnarray}
\dot\rho+3(p+\rho)H= -\frac{\dot\Lambda}{8\pi G}.
\end{eqnarray}

\section{Cosmological model}
Let us consider a relationship between the fluid pressure and
density of the physical system in the form of the following
barotropic equation of state
\begin{eqnarray}
p= \omega\rho
\end{eqnarray}
where $\omega$ is the equation of state parameter. In this
barotropic equation of state $\omega$ is assumed to be a
time-dependent quantity i.e. $\omega=\omega(t)$ . Actually, the
above equation of state parameter $\omega$, instead of being a
function of time, may also be function of scale factor or
redshift. However, sometimes it is convenient to consider $\omega$
as a constant quantity because current observational data has
limited power to distinguish between a time varying and constant
equation of state \cite{Kujat2002,Bartelmann2005}. In this
connection it may be useful to stress that equation (6) is
assigned to matter content of the Universe (with possible time
dependent $\omega <0$, which is not related with $\Lambda$).

Now, using equation (6) in (5) we get
\begin{eqnarray}
8\pi G\dot\rho+\dot\Lambda= -24\pi G(1+\omega)\rho H.
\end{eqnarray}

Again, differentiating (3) with respect to $t$ we get for a flat
Universe ($k=0$)
\begin{eqnarray}
4\pi G\rho=- \frac{\dot H}{1+\omega}.
\end{eqnarray}

It can be mentioned that equivalence of three phenomenological
$\Lambda$-models (viz., $\Lambda \sim (\dot a/a)^2$, $\Lambda \sim
\ddot a/a$ and $\Lambda \sim \rho$) have been studied in detail by
Ray et al. \cite{Ray2007} for constant $\omega$. So, it is
reasonable that similar type of variable-$\Lambda$ model may be
investigated with a variable $\omega$ for a deeper understanding
of both the accelerating and decelerating phase of the Universe.
Let us, therefore, use the {\it ansatz} $\dot\Lambda \propto H^3$,
so that
\begin{eqnarray}
\dot\Lambda= AH^3
\end{eqnarray}
where $A$ is a proportional constant. This {\it ansatz} with
negative $A$ can find realization in the approach of
self consistent inflation
\cite{Dymnikova:2001ga,Dymnikova:2001jy,Dymnikova:1998ps}, in
which time-variation of $\Lambda$ is determined by the rate of
Bose condensate evaporation \cite{Dymnikova:2001ga} with $\alpha \sim
(m/m_{Pl})^2$ (where $\alpha$ is the absolute value of negative $A$ and
$m$ is the mass of scalar field and $m_{Pl}$ is the Planck mass).

The physical nature of $\Lambda$ in this approach is related with
the energy density of scalar field, which has the amplitude beyond
the minimum of scalar field potential. If the mass of scalar field
$m$ strongly exceeds Hubble parameter $H$, $m \gg H$, coherent
field oscillations should start, being accompanied by the
condensate evaporation. The interaction of evaporated particles
with condensate damp coherent oscillations of scalar field and
keep the amplitude of this field beyond the minimum of its
potential. It makes the rate of dissipation of $\Lambda$ dependent
on the cross section of particle interaction with condensate. The
maximal estimation of this cross section leads
\cite{Dymnikova:2001ga,Dymnikova:2001jy} to the time variation of
$\Lambda$, given by Eq. (9).

Using equations (6), (8) and (9) we get from (4),
\begin{eqnarray}
\frac{2}{(1+\omega)H^3}\frac{d^2H}{dt^2}+\frac{6}{H^2}\frac{dH}{dt}=A.
\end{eqnarray}
If we put $dH/dt=P$, then equation (10) reduces to
\begin{eqnarray}
\frac{dP}{dH}+3(1+\omega)H= \frac{A(1+\omega)H^3}{2P}.
\end{eqnarray}

To arrive at any fruitful conclusion, let us now solve equation
(11) under the following specific assumption
\begin{eqnarray}
\omega(t)= -1 + \frac{2\tau P}{H}.
\end{eqnarray}
 where $\tau$ has dimension of time and is a parameter of our model.
Typically, the time scale $\tau$ has the physical meaning of
dissipation time scale for time varying $\Lambda$. Here $\tau$
comes in the picture due to the dimensional requirement. As stated
earlier, $\omega$ is the equation of state parameter which depends
upon time i.e. $\omega=\omega(t)$. This time-dependence may be
represented by functional relationship with cosmic scale factor
$a$ or cosmological redshift $z$. In connection to redshift this
dependence may be linear as $\omega(z) = \omega_o +
{\omega}^{\prime} z$ where ${\omega}^{\prime}= (d\omega/dz)_{z=0}$
\cite{Huterer2001,Weller2002} or may be of non-linear type as
$\omega(z) = \omega_o + {\omega}_1 z/(1+z)$
\cite{Polarski2001,Linder2003}. Now, following these redshift
parameterizations of two index pattern we can assume our above
equation of state parameter in the form $\omega(t) = \omega_o +
{\omega}_1 [1+(t/\tau)^2]/[1+(t/\tau)]$ with $ \omega_o =-1$ and
$\omega_1 = 2$. The form of this supposition, later on, will be
realized from the solution (18).

By the use of above supposition (12), equation (11) becomes
\begin{eqnarray}
\frac{dP}{dH}+6\tau P= A\tau H^2.
\end{eqnarray}

Therefore, from the master equation (13) we get the solution set
as
\begin{eqnarray}
a(t)= C_1e^{t/6\tau}\left(sec\frac{Bt}{\tau}\right)^{1/6B},
\end{eqnarray}
\begin{eqnarray}
H(t)= \frac{1}{6\tau}\left(1+tan \frac{Bt}{\tau}\right),
\end{eqnarray}
\begin{eqnarray}
\Lambda(t)= \frac{1}{6\tau^3}\left[\frac{\tau}{2} tan^2
\frac{Bt}{\tau}+2\tau log \left (sec \frac{Bt}{\tau}\right)+3\tau
tan \frac{Bt}{\tau}-2Bt\right],
\end{eqnarray}
\begin{eqnarray}
\rho(t)=- \frac{1}{48\pi G\tau^2}\left(1+tan
\frac{Bt}{\tau}\right),
\end{eqnarray}
\begin{eqnarray}
\omega (t)= -1+\frac{2Bsec^2\frac{Bt}{\tau}}{\left(1+tan
\frac{Bt}{\tau}\right)},
\end{eqnarray}
\begin{eqnarray}
p(t) = \frac{1}{48\pi G\tau^2}\left(1+tan \frac{Bt}{\tau}-2Bsec^2
\frac{Bt}{\tau}\right)
\end{eqnarray}
where $C_1$ is a constant and B=A/36.

It is interesting to note that, since the equation (15) contains
tangent function with a singularity at $\pi/2$, the model exhibits
the properties of an oscillating one. This particular point will
be demonstrated graphically in the next Section 4 for different
physical parameters.

\section{Physical features of the constants and parameters}

\subsection{Structure of $A$}
 As evident from the equation (9) $A$ is, by construction, dimensionless
while B has dimension of inverse time. Therefore, to clarify the
units in which one is to measure B we propose that the equation
(9) can be taken in the form $\dot \Lambda = A(t) H^3$ where the
constant of proportionality $A$ is now assumed as time-dependent
with the form $A(t) = A_0 + A_1 t^{-1}$. In view of this the
equation (9) can be considered now as the truncated linear form
with constant $A=A_0$ only.

\subsection{Nature of $B$}
For physically valid $H$ we should have $tan ({Bt}/{\tau})>-1$.
Again, from equation (18) it is clear that $\omega$ can be greater
or less than $-1$ according as $B>0$ or $B<0$. But, the awkward
case here is the negativity of $\rho$. In this connection it may
be mentioned that Ray and Bhadra~\cite{Ray2004a} also obtained
negative energy density by introducing a space-varying $\Lambda$
for a static charged anisotropic fluid source. In fact, according
to Cooperstock and Rosen~\cite{Cooperstock1989}, Bonnor and
Cooperstock~\cite{Bonnor1989} and Herrera and
Varela~\cite{Herrera1994}, within the framework of general theory
of relativity some negative mass density must be possessed by any
spherically symmetric distribution of charge. Ray and
Bhadra~\cite{Ray2004b} have demonstrated that, model constructed
within Einstein-Cartan theory can also contain some negative
matter-energy density. In the present work a negative density is
obtained for a dynamic, homogeneous and isotropic neutral fluid
with time dependent $\Lambda$. It is shown here that in $H^3$
model accelerated expansion of the Universe takes place at
negative energy density, but at positive pressure (for negative
$B$, we have negative energy density and negative $\omega$, giving
positive pressure). In fact, at negative $B$ time varying
cosmological term disappears at time scale $t \sim \tau/b =
36\tau/a$ (where $b$ is the absolute value of negative $B$),
therefore addition of dark matter will lead to matter dominance at
$t>\tau/b$.

This result is quite natural, since negative $B$ corresponds to
negative $A$, i.e. to negative time derivative for $\Lambda$ in
equation (9). The physical meaning of $\tau$ is straightforward
 for each particular realization of the considered scenario.
 For example in the approach \cite{Dymnikova:2001ga,Dymnikova:2001jy,Dymnikova:1998ps}
it has the meaning of timescale, at which Bose-Einstein
condensate, maintaining $\Lambda$, evaporates.

\subsection{Equation of state parameter $\omega$}
It is observed from the equation (18) that unless the second term
vanishes $\omega$ can not be negative as expected from the SN Ia
data~\cite{Knop2003} and SN Ia data with CMB anisotropy and
galaxy-cluster statistics~\cite{Tegmark2004b} in connection to
dark energy. In this regards we would like to mention here that
the physical significance of the negative density can be realized
if one remembers that in the present investigation the dark energy
is considered not through the equation of state (6) with negative
$\omega$ rather through the ansatz for $\Lambda$ where $\Lambda$
acts as one of the dark energy candidates. With this
consideration, $\Lambda$ makes a definite contribution and
ultimately the positive pressure is over-powered by $\Lambda$ to
make it negative and energy density positive so that the Universe
becomes accelerating. Since the expression of $\omega(t)$ chosen
here contains a term $(1 + t/\tau)$ in the denominator, the power
series expansion of it (if permissible) reveals that the
expression for $\omega(t)$ contains positive and negative terms
alternately. In that case, the Equation of State is an oscillating
one (see Figs. 1 \& 2).

\subsection{Deceleration parameter $q$}
We would like to consider now the equation (15) which yields the
expression for the deceleration parameter as $ q= -[1 + {6B sec^2
{\frac{Bt}{\tau}}}/(1 + tan{\frac{Bt}{\tau}})^2]$. It is clear
from this expression that if $B<0$, then $q$ can change its sign
depending on the value of the time dependent part. So,
decelerating-accelerating cosmic evolution can be found from the
present $H^3$ phenomenological $\Lambda$-CDM model. The expression
for the deceleration parameter $q$ involves tangent function. So,
like Eq. (15), it also has a singularity at $\pi/2$ (see Figs. 3,
4, 5 and at larger $t$ the parameter $q$ may be oscillating.

\section{Discussions and Conclusions}
The objective of this work is to observe the effect of a
time-dependent equation of state parameter on a dynamic $\Lambda$
model selected for dark energy investigation. Assuming $\dot
\Lambda\sim H^3$, expressions for time dependent equation of state
parameter and matter density have been derived. This assumption
can find physical justification through the model of Bose-Einstein
condensate evaporation~\cite{Dymnikova:2001ga,Dymnikova:2001jy}.
It is also to be noted that the change of sign of the deceleration
parameter has been achieved under this special assumption. Our
solutions indicate very nontrivial features of cosmological
parameters (see Figs. 1 - 11). They are mostly oscillating and
contain singularities. In particular, singular and oscillating
nature of $\omega(t)$ is demonstrated for large $t$ on Fig. 2

Moreover, from the solution set it is clear that the expression
for density comes out to be a negative quantity with a positive
pressure counter part. Actually, here what is shown is a possible
cosmological solution with negative energy density. The mechanism
of the accelerated expansion of the present Universe due to this
positive pressure and negative density is not yet understood
properly. In general, contemporary literature (see e.g.
\cite{Ray2007} and references therein) suggest that via
Cosmological parameter $\Lambda$-dark energy acts as the role for
the repulsive pressure which is responsible for the accelerated
expansion. Here, in our case, negative energy density seems to
possess the same role of repulsive gravitation. Perhaps it's role
could be understood in a model with dark matter in the presence of
$\Lambda$ where dark matter will be associated with repulsive
nature due to negative density \cite{Goodman2000}.

The main idea of the result is that there is an acceleration due
to negative energy density, but at positive pressure. Though
physically it is an awkward situation but not at all unavailable
in the literature (as mentioned earlier in the introduction) and
also quantum field theory admits it. However in cosmology, we
don't see negative energy density phenomena very often. This is
because there may be some mechanism restricting negative energies
or their interactions with ordinary fields. The idea of
anti-gravity (i.e. gravitational repulsion between matter and
antimatter) originates from this kind of negative energy
\cite{Hossenfelder2006}. Due to symmetry, each gravitating
standard model particle corresponds to an anti-gravitating
particle. It is thought that this anti-gravitating particle
cancels gravitating particle's contribution to the vacuum energy
and hence provides a mechanism for smoothing out of the
cosmological constant puzzle \cite{Quiros2004}.

We would also like to point out further that-

1. The present work has been done by keeping the gravitational
constant $G$ as a constant. Therefore, a future work can be
carried out along the line of this model with $G$ as a variable.

2. Unless we study {\it dark matter} + $\Lambda$, we can not say
anything conclusive about $\Lambda$-CDM cosmology.

3. It is also interesting to consider early Universe for $\Lambda$
+ {\it radiation}, or more general $\Lambda$ + {\it radiation} +
UDM, and then to study physics of such Universe and possible
observational constraint.

4. It is not very easy to calculate the present age of the
Universe from the expressions of various parameters derived in the
present work. However, numerical methods may be helpful in
calculating the present age from Eq. (15) by using different
values of $H_0$ given in  \cite{Ray2007}. Since the purpose of the
present paper is different, so that attempt has not been made
here.

\section*{Acknowledgments}
One of the authors (SR) is thankful to the authority of
Inter-University Centre for Astronomy and Astrophysics, Pune,
India for providing him Associateship programme under which a part
of this work was carried out. We all are thankful to the referee
for valuable suggestions which have enabled us to improve the
manuscript substantially.




{}

\pagebreak

\begin{figure*}
\begin{center}
\vspace{0.5cm}
\includegraphics[width=0.6\textwidth]{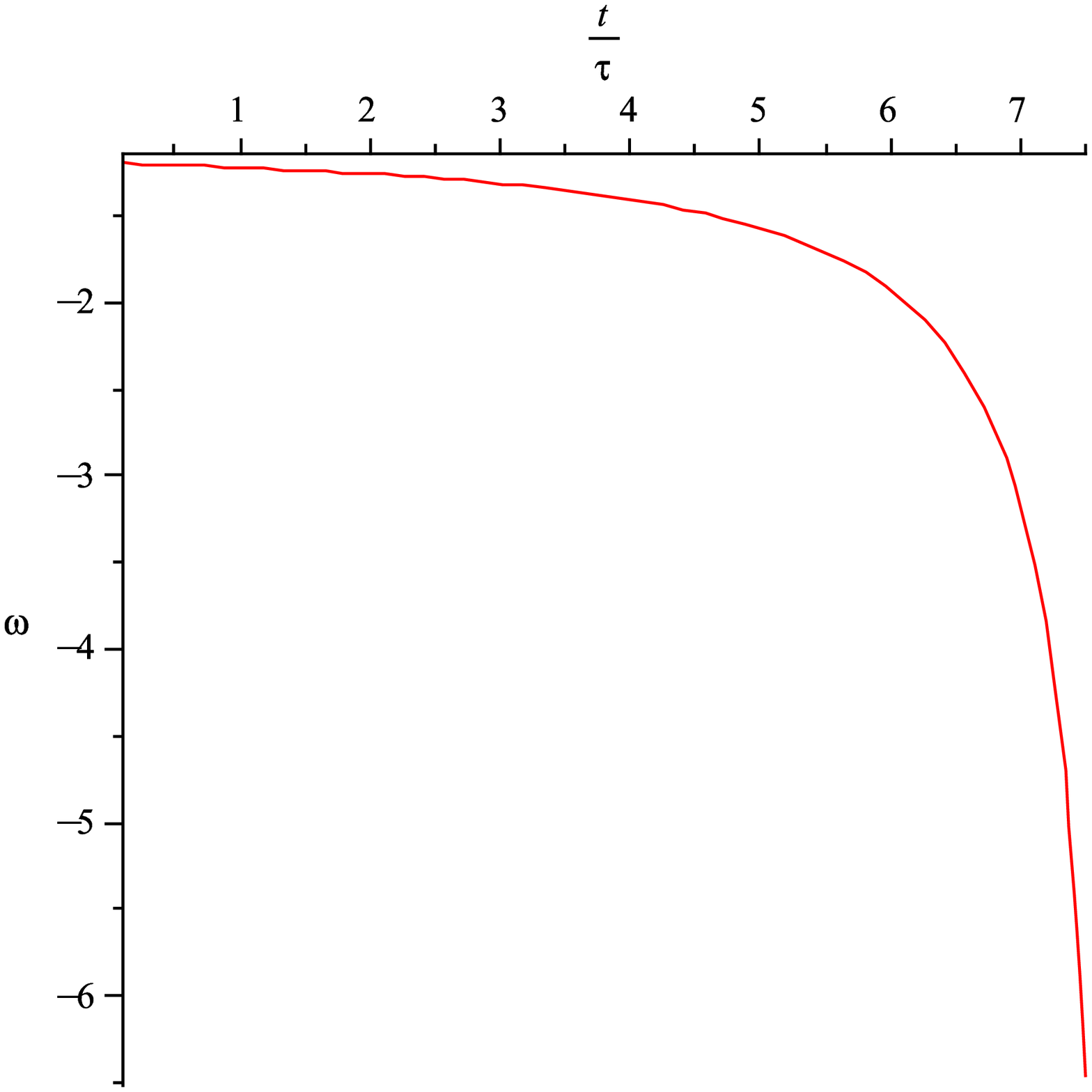}
\caption{Variation of $\omega$ with small values of time $t$.}
 \label{fig1}
\end{center}
\end{figure*}

\begin{figure*}
\begin{center}
\vspace{0.5cm}
\includegraphics[width=0.6\textwidth]{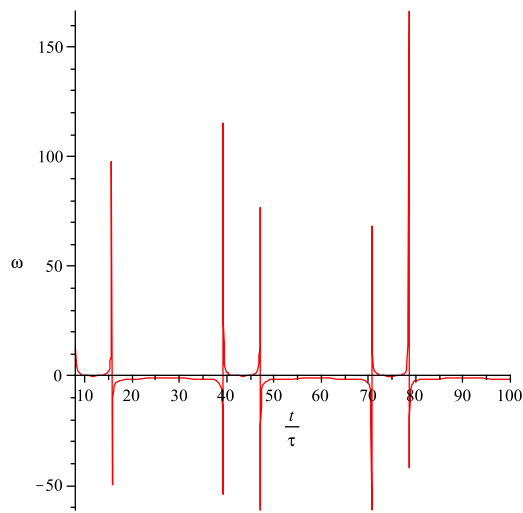}
\caption{Variation of $\omega$ with large values of time $t$.}
 \label{fig2}
\end{center}
\end{figure*}

\begin{figure*}
\begin{center}
\vspace{0.5cm} \includegraphics[width=0.6\textwidth]{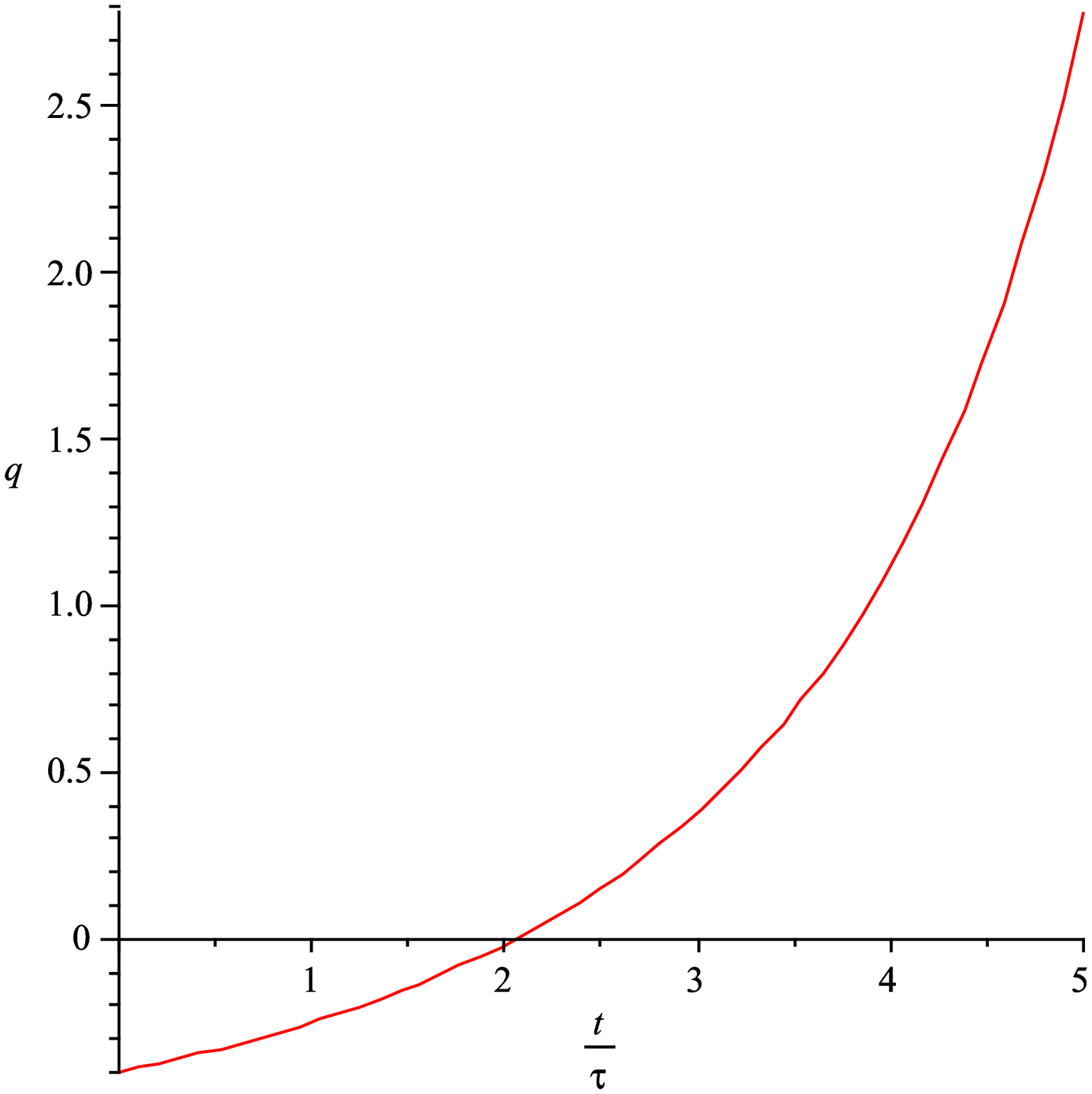}
\caption{Variation of deceleration parameter $q$ with time $t$.}
 \label{fig3}
\end{center}
\end{figure*}

\begin{figure*}
\begin{center}
\vspace{0.5cm}
\includegraphics[width=0.6\textwidth]{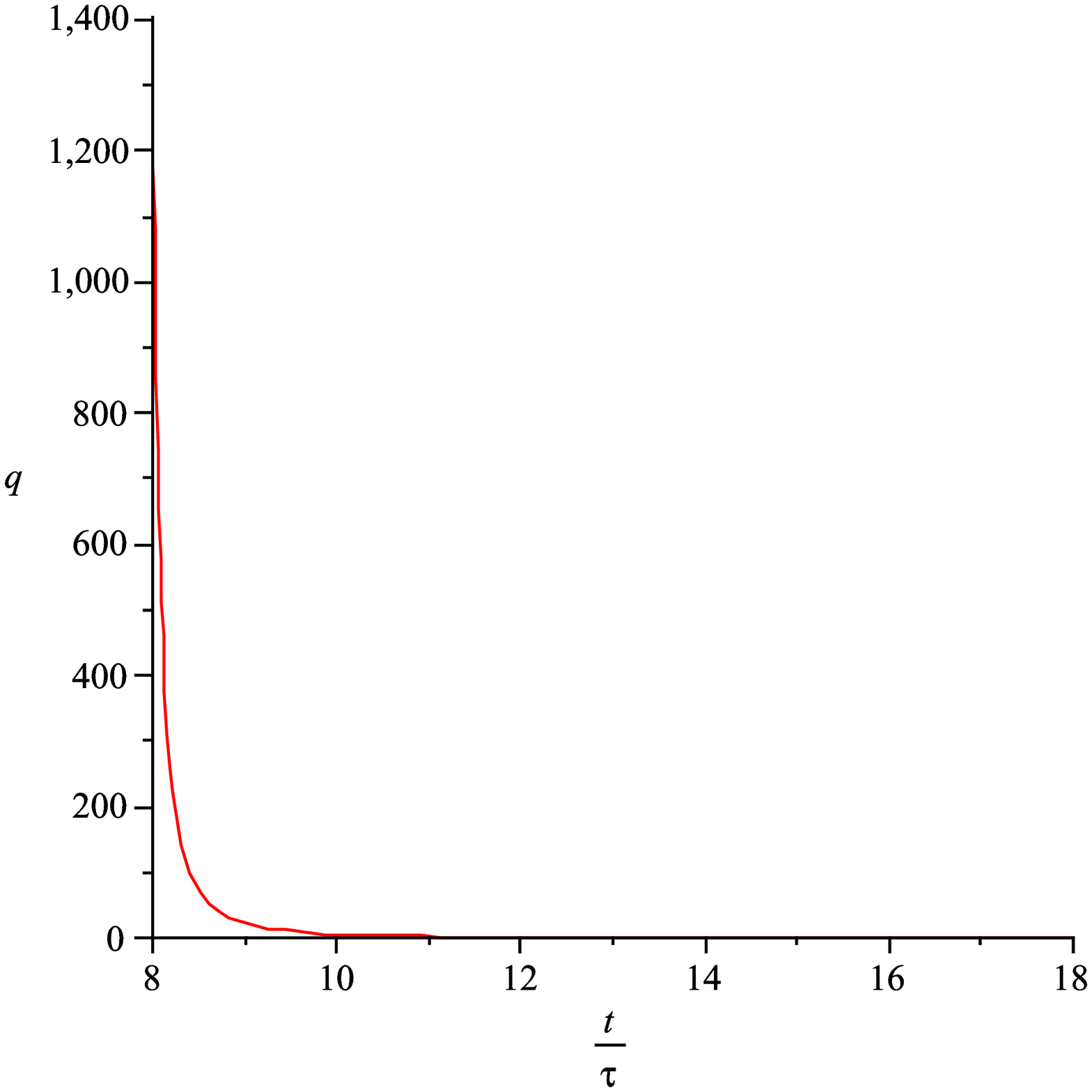}
\caption{Variation of deceleration parameter $q$ with large values
of time $t$.}
 \label{fig4}
\end{center}
\end{figure*}

\begin{figure*}
\begin{center}
\vspace{0.5cm}
\includegraphics[width=0.6\textwidth]{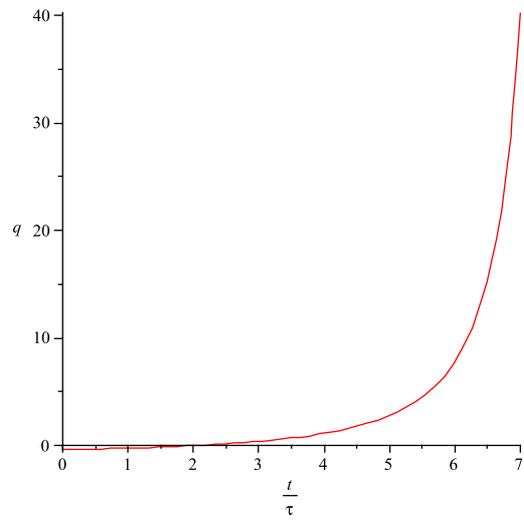}
\caption{Variation of deceleration parameter $q$ with some
specified values of time $t$.}
 \label{fig5}
\end{center}
\end{figure*}

\begin{figure*}
\begin{center}
\vspace{0.5cm} \includegraphics[width=0.6\textwidth]{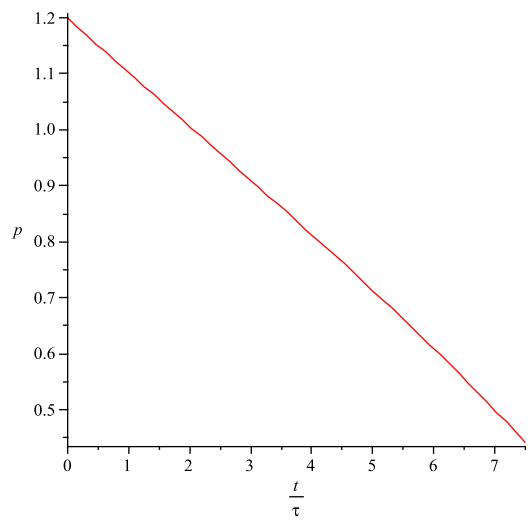}
\caption{Variation of pressure $p$ with small values of time $t$.}
 \label{fig6}
\end{center}
\end{figure*}

\begin{figure*}
\begin{center}
\vspace{0.5cm}
\includegraphics[width=0.6\textwidth]{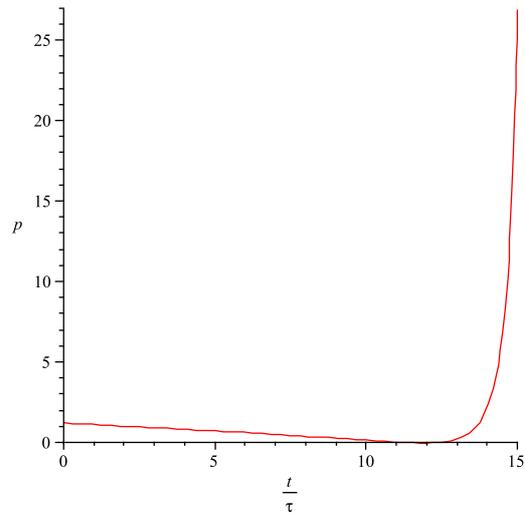}
\caption{Variation of pressure $p$ with large values of time $t$.}
 \label{fig7}
\end{center}
\end{figure*}

\begin{figure*}
\begin{center}
\vspace{0.5cm} \includegraphics[width=0.6\textwidth]{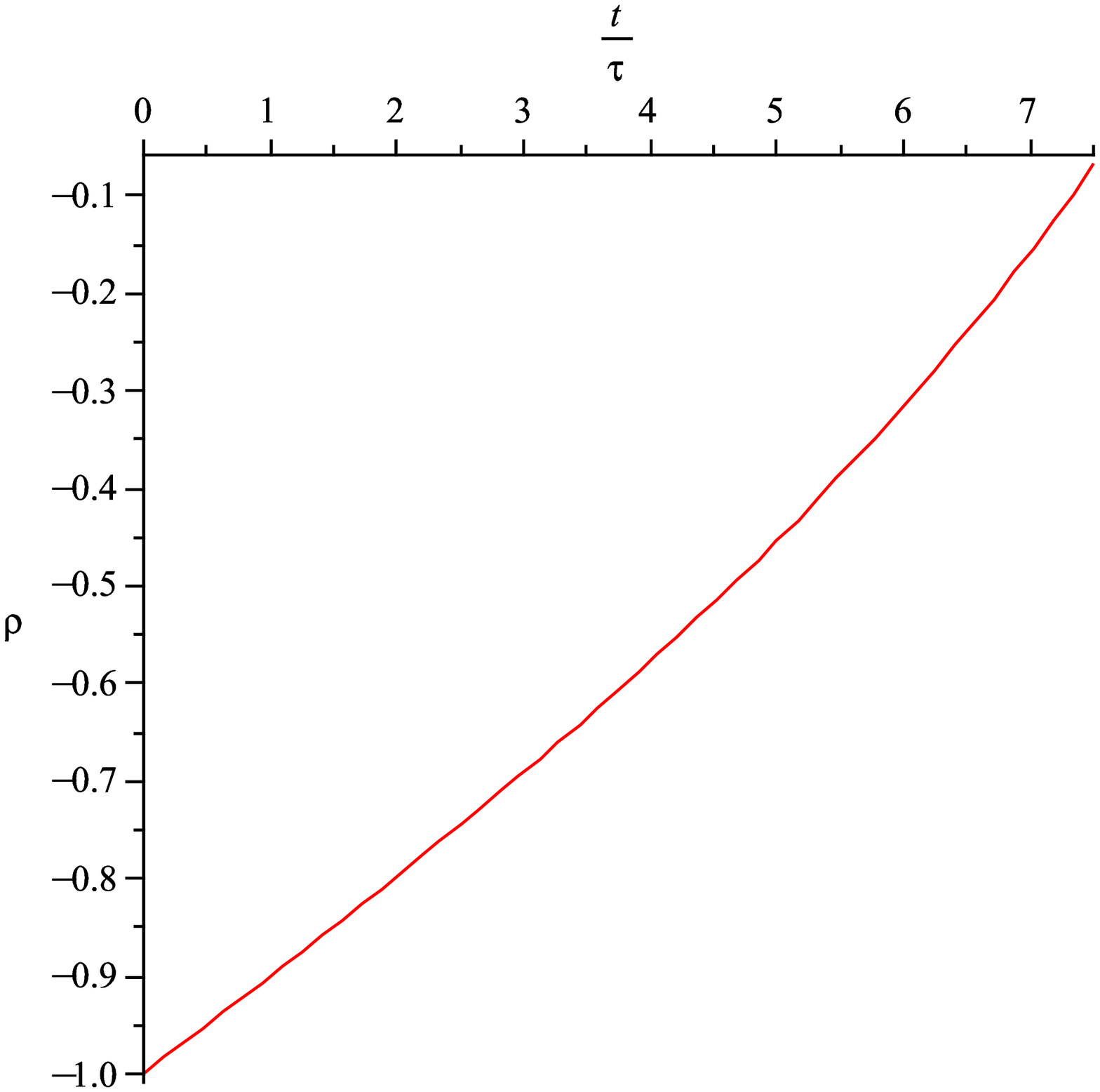}
\caption{Variation of matter-energy density $\rho$ with time $t$.}
 \label{fig8}
\end{center}
\end{figure*}

\begin{figure*}
\begin{center}
\vspace{0.5cm}
\includegraphics[width=0.6\textwidth]{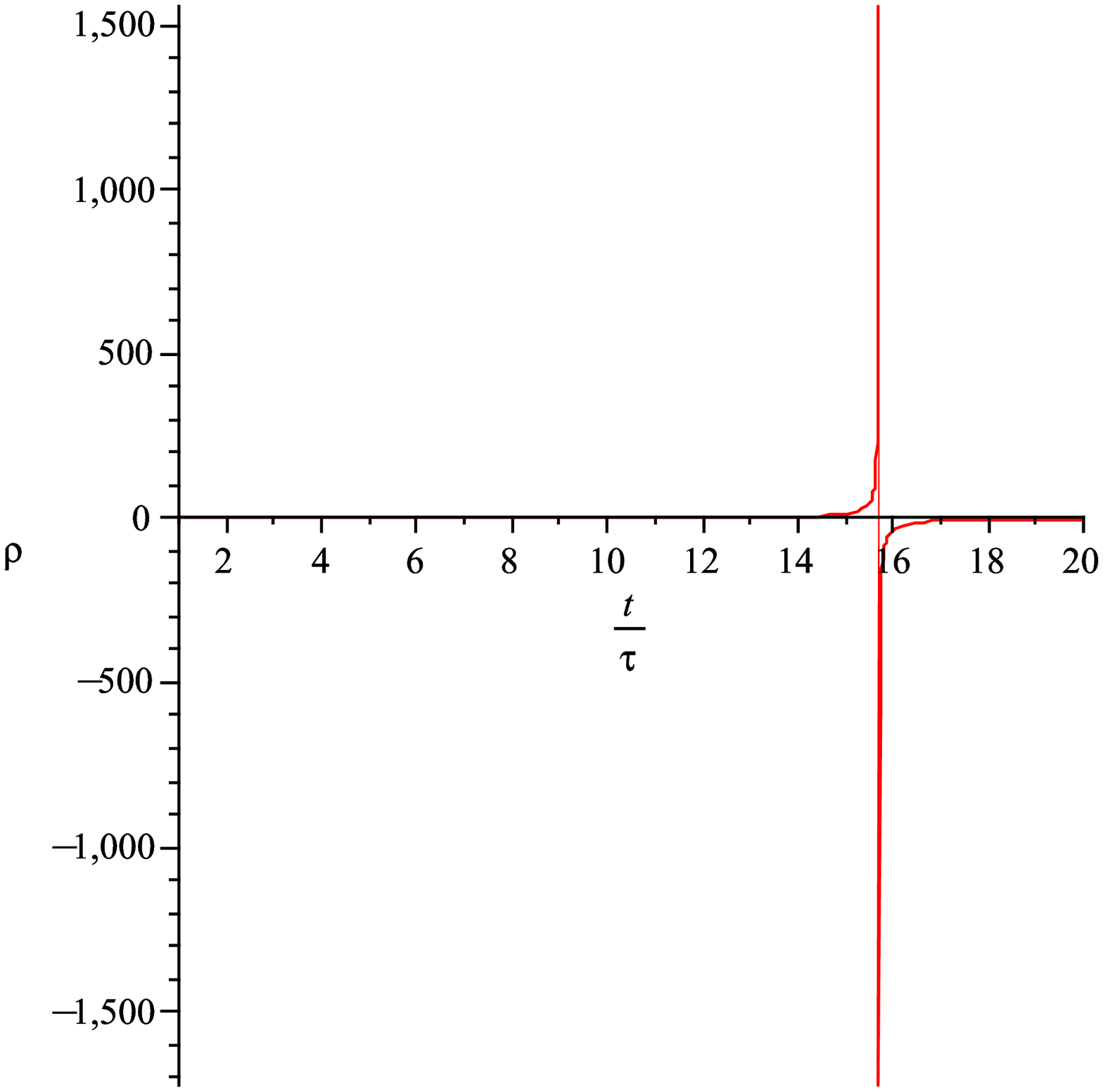}
\caption{Variation of matter-energy density $\rho$ with large
values of time $t$.}
 \label{fig9}
\end{center}
\end{figure*}

\begin{figure*}
\begin{center}
\vspace{0.5cm}
\includegraphics[width=0.6\textwidth]{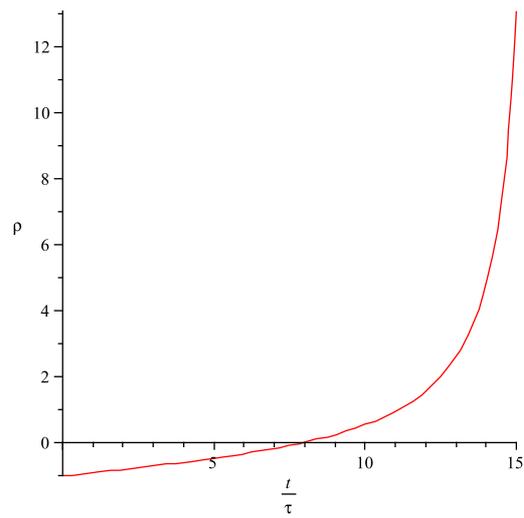}
\caption{Variation of matter-energy density $\rho$ with some
specified values of time $t$.}
 \label{fig10}
\end{center}
\end{figure*}

\begin{figure*}
\begin{center}
\vspace{0.5cm}
\includegraphics[width=0.6\textwidth]{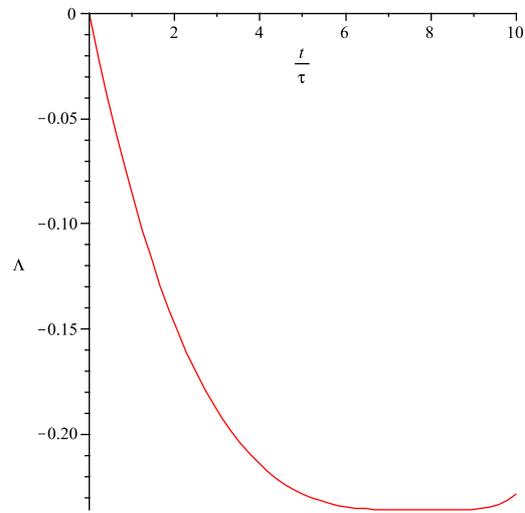}
\caption{Variation of cosmological parameter $\Lambda$ with time
$t$.}
 \label{fig11}
\end{center}
\end{figure*}

\end{document}